\pdfoutput=1

\documentclass[fleqn,usenatbib]{mnras}

\usepackage{subfigure,epsfig,url,bm,amsmath,cases,color,soul,amssymb}

\usepackage{comment}

\title[Kuiper Belt Binary Evolution]{Evolution of Primordial Kuiper Belt Binaries Through a Giant Planet Instability}

\author[Stone \& Kaib.]{
Lukas R. Stone,$^{1}$\thanks{E-mail: lrs5694@psu.edu}
Nathan A. Kaib,$^{1}$
\\
$^{1}$HL Dodge Department of Physics \& Astronomy, University of Oklahoma, Norman, OK 73019, USA\\
}

\date{Accepted 2021 April 28. Received 2021 April 26; in original form 2021 February 22}

\begin{document}
\label{firstpage}
\pagerange{\pageref{firstpage}--\pageref{lastpage}}
\maketitle

\begin{abstract}
The non-resonant Kuiper belt objects (KBOs) between the 3:2 and 2:1 Neptunian mean motion resonances can be largely divided between a cold classical belt (CCB) and a hot classical belt (HCB). A notable difference between these two subpopulations is the prevalence of widely spaced, equal-mass binaries in the CCB and a much smaller but non-zero number in the HCB. The primary reason for this difference in binary rate remains unclear. Here using N-body simulations we examine whether close encounters with the giant planets during an early outer solar system instability may have disrupted primordial Kuiper belt binaries that existed within the primordial Kuiper belt before they attained HCB orbits. We find that such encounters are very effective at disrupting binaries down to separations of $\sim$1\% of their Hill radius (as measured in the modern Kuiper belt), potentially explaining the paucity of widely spaced, equal mass binaries in the modern HCB. Moreover, we find that the widest binaries observed in the modern HCB are quite unlikely to survive planetary encounters, but these same planetary encounters can widen a small subset of tighter binaries to give rise to the small population of very wide binaries seen in today's HCB. 
\end{abstract}


\section{Introduction}

The Kuiper belt is a collection of small icy bodies located past Neptune at 30 AU and extends outward beyond 50 AU. The Kuiper belt is classified into subpopulations based on dynamical considerations: The resonant population, scattered disk, detached objects and the classical belt \citep{gladman2008nomenclature}. The classical belt is defined to be Kuiper belt Objects (KBOs) far enough from Neptune to be stable but with semimajor axes exterior to the 3:2 resonance with Neptune and closer than the 2:1 resonance. Based on the observed KBOs, the classical belt can be described further by a bimodal distribution of bodies: The Cold Classical Kuiper belt (CCB) and the Hot Classical Kuiper belt (HCB) \citep[e.g.][]{gulbis2010unbiased,brown2012compositions}.

The CCB and HCB differ in both physical properties and orbital parameters. Comparing classical belt objects with low orbital inclinations ($i\lesssim5^{\circ}$) to classical belt objects with higher inclinations ($i\gtrsim5^{\circ}$) reveals key physical differences. (It should be noted that an inclination division is imperfect as the inclination distribution near $i=5^{\circ}$ suggests a significant amount of mixing between the CCB and HCB \citep{volk2011inclination}). CCB objects tend to have larger albedos \citep{brucker2009high}, redder colors \citep{tegler2000extremely}, smaller sizes \citep{levstern01}, and lower orbital eccentricities \citep{petit11}. Moreover, a notable percentage of CCB objects exist in binary systems, with a sub-population of ultra wide binaries (UWB) in which binary companions are separated by at least 7$\%$ of their Hill radius \citep{noll2008solar,noll2008evidence}. Meanwhile, the HCB appears to be nearly devoid of equal-mass binary systems analogous to those seen in the CCB, yet a handful, at the level of 1--2\% of the HCB population, exist \citep{nesvorny2019binary}. The physical differences between the CCB and HCB suggest that these two groups of bodies formed in different regions of the solar system and underwent separate evolutionary paths. It is commonly hypothesized that the CCB formed in situ at its current position ($>40$ AU), while the HCB formed closer to the sun ($25-30$ AU) and was superimposed onto the CCB by an outward migrating Neptune \citep[e.g.][]{nesvorny2018dynamical}. 

A solar system evolutionary framework known as the Nice Model posited that the early solar system underwent an instability which resulted in Uranus and Neptune migrating outward toward their current positions (\cite{tsiganis2005origin}). As Neptune migrated outward, it plowed through a massive ($\sim$10--30 M$_{\oplus}$) primordial Kuiper belt orbiting between $\sim$25--30 AU, capturing many bodies in resonances. Shown by \citet{malhotra1993origin}, bodies caught in resonance with an outwardly migrating Neptune will be carried along and have their eccentricities and inclinations increased, and some of these objects may subsequently escape resonance into HCB-like orbits. Therefore, a migrating Neptune may offer an explanation for the observed differences between an in-situ CCB and a transported HCB. This transport scenario can explain the inclination and eccentricity differences between the HCB and CCB, and their different formation locations offer a plausible explanation for the color, size, and albedo differences. Moreover, it may explain the paucity of binary objects among the HCB. Within this framework, one possibility is that primordial belt objects simply did not often form in binary configurations, but another possibility is that primordial belt binaries were as prevalent as modern CCB binaries, but the primordial binaries were dissociated before reaching the modern HCB. 

Prior to HCB formation, binaries can be disrupted in a couple different ways: impulses from impacting primordial belt objects could dissociate such binaries if the belt was long-lived ($\gtrsim$100 Myrs), or close encounters with the giant planets could dissociate companions during the belt's dispersal. Both of these processes have been studied previously, and it has been found that planetary encounters are quite efficient at disrupting primordial binaries. However, some of this work has focused on objects that could have been implanted into the CCB \citep{parker2010destruction}, while others have studied very specific orbital evolution scenarios for the giant planets \citep{nesvorny2019binary}. 

Here we model the dynamical fates of primordial Kuiper belt binaries during their encounters with planets under a variety of less constrained outer solar system instabilities. We expect that none of these systems are a perfect match to the actual solar system but that the full range of behavior they exhibit brackets that which occurred in the real solar system. Our work is structured as follows: In Section 2, we discuss the setup and pipeline of our N-body simulations. In Section 3, we discuss the results of these simulations. In Section 4, we discuss the implications and significance of our results. Finally, in Section 5, we summarize our work.

\section{Numerical Methods}
 
 While previous studies have modeled the dynamical fates of primordial KBO binaries entering the HCB via planetary encounters, they use a forced migration of the planets through a disk of massless test particles representing primordial KBOs, testing a single, unique planetary migration scenario and instability \citep{nesvorny2019binary}. Here we seek to model these encounters as they occur within a system undergoing a full planetary instability in which the planetary orbital evolution is caused by gravitational interactions with massive KBOs. This strategy will likely result in a different distribution of planetary encounters since planetary instabilities are highly chaotic and there will be feedback between the rate of planetary orbital evolution and the dynamical excitation of the primordial Kuiper belt. 
 
 To model the dynamical effects that planetary encounters have on binaries of the primordial Kuiper belt, we must first simulate the planet-driven dispersal of the primordial Kuiper belt and formation of the modern HCB. Based off numerous previous works, we assume that Neptune's migration through the massive primordial belt, interrupted by a giant planet orbital instability, is the mechanism for primordial belt dispersal and the formation of the modern HCB \citep{malhotra1993origin, nes15}. To initiate this process, we place Jupiter, Saturn, and three 15 M$_{\oplus}$ ice giants in a 3:2, 3:2, 2:1, 3:2 resonant chain surrounded by a 20 M$_{\oplus}$ belt of 1000 planetesimals orbiting between 21.4 and 30 AU (see \citet{nesmorb12} and \citet{ray20} for additional details). Because each planetary orbital instability is unique and we wish to study KBO binaries in a variety of different instabilities, we design 100 such systems (each with uniquely sampled disk particle orbits) and integrate them for 1 Gyr with the MERCURY hybrid integrator \citep{chambers1999hybrid}. Virtually all of our systems pass through instability early in their integrations, but only 18 collapse to 4-planet systems with Jupiter, Saturn, and two ice giants in the proper order. We take this ordering of the planets as a very broad criterion for ``solar system analogs,'' since the inclusion of more demanding constraints quickly drives down the number of usable systems and may not have a dramatic impact on the very broad characteristics of Kuiper belt architecture. 
 
 For each of the 18 Nice model simulations, general bounds are set to define surviving particles residing in the modern HCB. Any particle at the end of a simulation located between the 3:2 and 2:1 resonance with Neptune ($\sim$39--48 AU) with a perihelion larger than 35 AU is taken to be a hot KBO. Because the trapping efficiency of the modern Kuiper belt is so low ($\sim$10$^{-3}$), not all of our 18 systems finish with hot belt particles. Only 11 systems are successful in creating between 1--4 KBOs. This results in a total number of 22 KBOs to simulate as equal mass binaries, running their entire close planetary encounter histories.
 
 In our Nice model simulations, every time an encounter within a planetary Hill radius occurs between a planetesimal and a planet, the two bodies' velocities and positions are saved as well as the time of the encounter into close encounter history data files. Using the close encounter histories between the planets and our KBO particles, we re-simulate these encounters, replacing the future HCB member with hypothetical binaries of varying mass and separation to study and assess their survival through HCB formation. For each close encounter simulation, conservation of energy and angular momentum in conjunction with the original Nice model simulations' close encounter data gives each encounter's initial planetocentric velocity ($V_{\infty}$) and impact parameter ($b$). Then the encountering planet of interest is situated at the origin, while the encountering planetesimal is now replaced with a hypothetical binary whose center of mass (COM) is positioned at $b$ and $V_{\infty}$. Then we use MERCURY's Bulirsch-Stoer integrator to simulate the binary's evolution through this individual encounter. The binary's COM initially starts at 5 Hill radii away from the planet of interest (where gravitational focusing has yet to become significant). The system is then integrated through the planetary encounter until the planet-binary distance again exceeds 5 Hill radii. The post-encounter binary orbit (binary semimajor axis, eccentricity, inclination, etc.) is then used as the pre-encounter binary configuration of each subsequent encounter. With this pipeline, we can assess how different hypothetical binaries are affected by the planetary encounters experienced by HCBs. 

Our simulations employ three different binary masses: (1) Binaries with a total mass of the UWB QW$_{322}$, or $2.2077 \times 10^{21}$ g \citep{parker2011characterization}. (2) Binaries with a total mass of an order of magnitude larger than QW$_{322}$, or $2.2077 \times 10^{22}$ g, and (3) Binaries with a total mass of an order of magnitude smaller than QW$_{322}$, or $2.2077 \times 10^{20}$ g. For all three mass values a total of five initial binary separations were modeled: 1$\%$, 3$\%$, 5$\%$, 7.5$\%$, and 10$\%$ of the binaries' Hill radius at 44 AU (the approximate middle of the modern classical Kuiper belt). A summary of our binary parameters are listed in Table 1. 
 
 \begin{table}
 \centering
 \begin{tabular}{lcr}
 \hline
 Initial Separation & $a_B/R_B$\\
   ($\%$ Hill radius*) & \\
 \hline
  1 & 46.5\\
  3 & 139.4\\
  5 & 232.4\\
  7.5 & 348.6\\
  10 & 464.8\\
  \hline
  \end{tabular}
 \caption{Simulated binary separations. Each separation employs three different binary masses: "Small" ($2.2077 \times 10^{20}$ g), "QW322" ($2.2077 \times 10^{21}$ g), and "Large" ($2.2077 \times 10^{22}$ g). 
 *At 44 AU}
 \end{table}
 
Motivated by the results of this initial set of simulations, we also run a follow-on set of simulations. These employ the same planetary encounters as the first set but focus in on a smaller range of binary separations. Unlike the first set, we parameterize the separations in terms of the ratio binary semimajor axis ($a_B$) to effective binary radius ($R_B$), or the cube root of the sum of the two companion's radii cubed ($(R_{1}^{3} + R_{2}^{3})^{1/3}$), where we assume that our binary objects have a bulk density of 0.5 g/cm$^3$. This second set of simulations has 360 different binaries with $a_B/R_B$ randomly sampled between 30 and 100. The total masses of the binaries used in this second set of simulations have the same three different mass values as the first set, with 120 binaries in each mass bin.
 
\section{Results}

\begin{figure}
\includegraphics[scale=0.24]{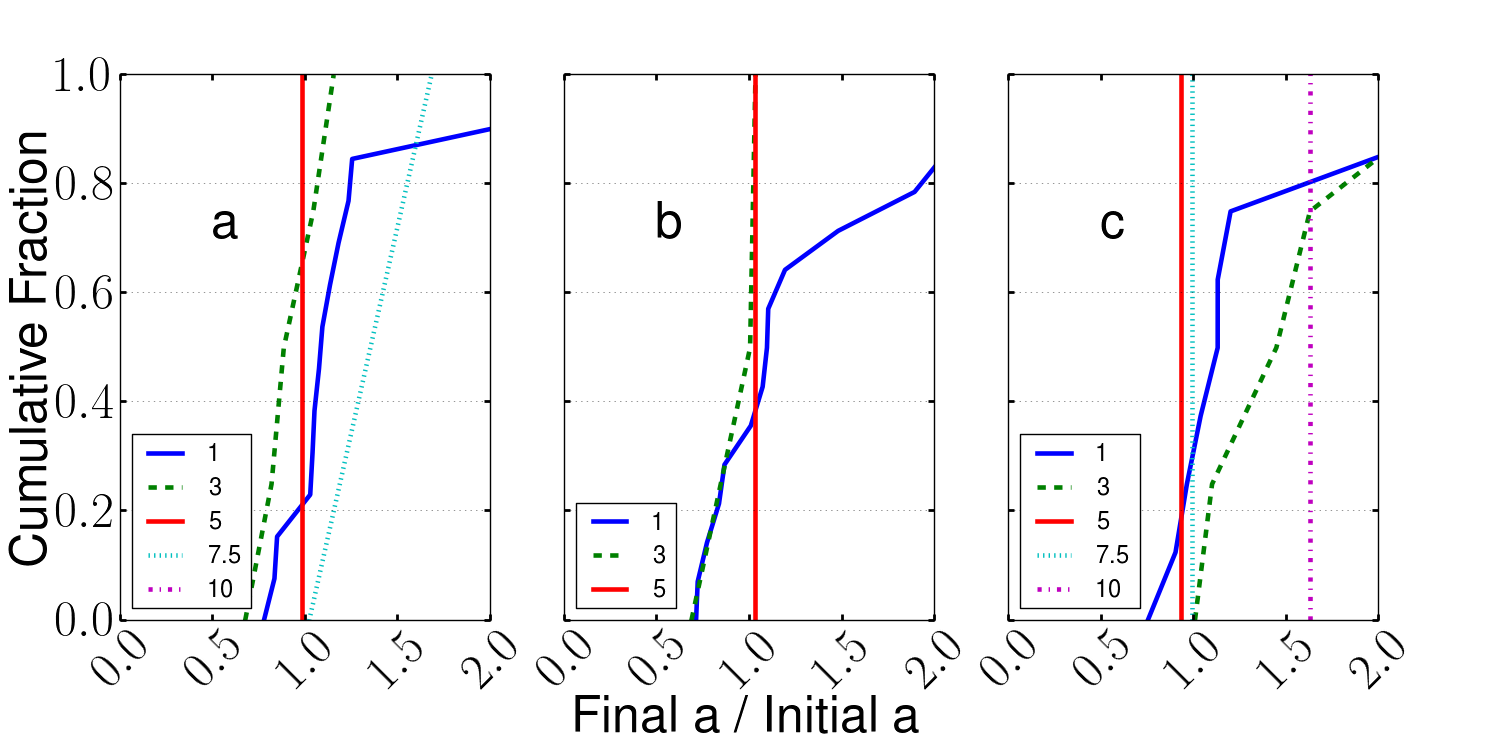}
\caption{The cumulative distributions of the final semimajor axis divided by the initial semimajor axis for all initial separations of surviving binaries. Plot (a) binaries have a total mass of an order of magnitude larger than QW322, Plot (b) binaries have a total mass of QW322, and Plot (c) binaries have a total mass of an order of magnitude smaller. NOTE: vertical lines are the cases in which only one binary survived.} 
\label{fig:SEMI_SUB}
\end{figure}

The purpose of these simulations is to assess the effect of close planetary encounters on potential binaries in the primordial KB. Fig. \ref{fig:SEMI_SUB} shows the distributions of the final-to-inital semimajor axis ratio for all three masses and initial separations for the binaries simulated. Plot (a) corresponds to the large mass ($2.2077 \times 10^{22}$ g), plot (b) corresponds to the total mass of the binary QW$_{322}$ ($2.2077 \times 10^{21}$ g, and plot (c) corresponds to the small mass ($2.2077 \times 10^{20}$ g). 

These results show that the total mass of the binary has little effect on the orbital evolution of the binaries. This is consistent with the results of \citet{nesvorny2019binary}, who find that the ratio of binary semimajor axis ($a_B$) to effective binary radius ($R_B$) is the parameter that governs binary survival through planetary encounters. We also find this to be the case. As seen in Table 2, there is a clear hierarchy associated with initial separation. Table 2 shows that our tightest binaries ($a_B/R_B \simeq 50$) have the greatest probability ($\sim$50\%) of surviving all the planetary encounters. This is at the high range of survival fractions that \citet{nesvorny2019binary} find, which we attribute to our greater diversity of planetary encounters owing to our multiple simulations of planetary instabilities. Meanwhile, any binary with an initial separation of $a_B/R_B \gtrsim 230$ has survival chances of $\sim$4\% or less. There is also a greater chance for tighter binaries to collide. This occurs when their eccentricity becomes large. This is also possible among our wider binaries, but collisions are rarer in this regime, as most of these binaries require even more extreme eccentricities and they also quickly become unbound before orbital eccentricity is allowed to evolve dramatically. 

A consequence of the results shown in Table 2 is that for most of the initial separations tested in our first set of simulations, little to no binaries survive their planetary encounters. Except for our tightest initial binary separations tested, the vast majority of the binaries are disrupted either through the binary components becoming unbound or, in a minority of instances, the binary components colliding. However, we can study the orbital evolution of the small percentage of binaries that do survive. 

The few surviving wider initial binaries typically have significant changes in eccentricity and inclination, as seen in Table 2, which is expected given how sensitive these fragile systems are to perturbation. Somewhat more surprisingly, the binaries initialized with a wider separation, (excluding the single 10$\%$ separation binary in plot c of Fig. \ref{fig:SEMI_SUB}) do not significantly evolve in semimajor axis. In these systems, the survivors represent a "lucky few" in which the set of planetary encounters they undergo lead to little energy change, whereas the ``unlucky'' binaries are rapidly dissociated by even modest planetary encounters. 

For our tightest initial binary separations (1$\%$ Hill radius separation), the results of Fig. \ref{fig:SEMI_SUB} often show a significant evolution in the binary semimajor axis. While it is possible for the planetary encounters to remove energy from systems, this effect is not dramatic, as illustrated by the small leftward tail of the cumulative histograms in Fig. \ref{fig:SEMI_SUB}. On the other hand, for all three masses tested, the tightest initial binaries have a large rightward tail seen in Fig. \ref{fig:SEMI_SUB} (a)-(c), indicating that a significant number of the surviving binaries had their semimajor axes inflated from  planetary encounters. In fact, 10--20\% of these binaries evolve to final separations that are at least double their initial separations, or $a_B/R_B \gtrsim 100$. 

\section{Discussion}

\begin{table*}
\centering
\begin{tabular}{lccccccr}
\hline
 Initial Separation/Mass & $\%$ Surviving & $\%$ Unbound & $\%$ Collided & Median Final Ecc. & Median $\Delta$ Inc.($^\circ$)\\
\hline
 
 1$\%$ Small & 40.90 & 45.45 & 13.63 & 0.172 & 8.935\\
 1$\%$ QW322 & 68.16 & 31.81 & 0.0 & 0.190 & 14.927\\
 1$\%$ Large & 63.63 & 31.81 & 4.54 & 0.287 & 16.083\\
 3$\%$ Small & 22.74 & 63.63 & 13.63 & 0.556 & 21.431\\
 3$\%$ QW322 & 13.63 & 77.27 & 9.09 & 0.102 & 23.031\\
 3$\%$ Large & 22.72 & 68.18 & 9.09 & 0.626 & 55.093\\
 5$\%$ Small & 4.54 & 90.90 & 4.54 & 0.029 & 25.719\\
 5$\%$ QW322 & 4.54 & 90.90 & 4.54 & 0.067 & 30.350\\
 5$\%$ Large & 4.54 & 81.81 & 13.63 & 0.075 & 15.186\\
 7.5$\%$ Small & 4.54 & 90.90 & 4.54 & 0.150 & 54.092\\
 7.5$\%$ QW322 & 0.0  & 95.45 & 4.54 & N/A & N/A\\
 7.5$\%$ Large & 9.09 & 90.90 & 0.0 & 0.604 & 90.600\\
 10$\%$ Small  & 4.54 & 95.45 & 0.0 & 0.429 & 26.496\\
 10$\%$ QW322  & 0.0  & 100.0 & 0.0 & N/A & N/A\\
 10$\%$ Large  & 4.54 &  95.45& 0.0 & 0.655 & 70.898\\
 \hline
\end{tabular}
 \caption{The percentage of surviving binaries, percent unbound, percent collided, median eccentricity and median delta inclination for all masses and initial separations investigated. Note: Small mass corresponds to $2.2077 \times 10^{20} g$, QW$_{322}$ corresponds to $2.2077 \times 10^{21} g$, and Large corresponds to $2.2077 \times 10^{22} g$. }
 \end{table*}
 
The survival results of our first set of simulations are largely in line with \citet{nesvorny2019binary} who find modest survival rates (10--50\%) for binaries with initial separations of $30 < a_B/R_B < 100$ and very small (0--5\%) survival rates for $a_B/R_B > 100$. They point out that the existence of a few very wide ($a_B/R_B > 100$) binaries within the HCB may require that a very large fraction of primordial KBOs initially belonged to such binaries, perhaps $\sim$40\%. This is much higher than the required 5--10\% fraction of slightly tighter ($30 < a_B/R_B < 100$) binaries in the primordial belt necessary to explain the modern population of these binaries in the HCB \citep{nesvorny2019binary}. 

However, our first set of simulations also show that planetary encounters transform a subset of binaries with initial separations of $a_B/R_B \simeq 50$ into very wide binaries with $a_B/R_B \gtrsim 100$. These results suggest that this transformed fraction may be up to $ \sim$10\% of the original $30 < a_B/R_B < 100$ population. This effectively means that if we assume that 5--10\% of primordial belt objects were in the form of binaries with $30 < a_B/R_B < 100$, then 0.5--1\% of modern HCBs should be binaries with $a_B/R_B > 100$. This lessens, and may altogether avoid, the high primordial fractions of very wide binaries implied in \citet{nesvorny2019binary}. 

\begin{figure}
\includegraphics[scale=0.39]{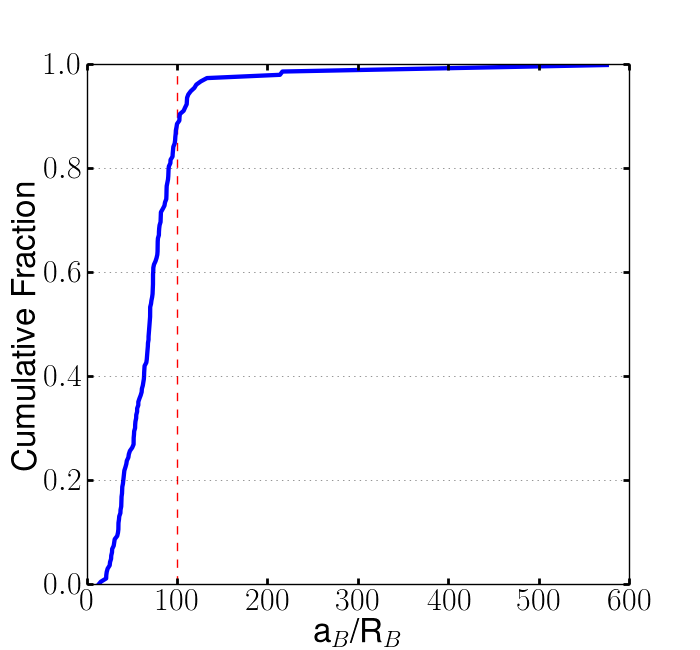}
\caption{The cumulative histogram of the final ratio of binary semimajor axes to binary radii ($a_B/R_B$) for binaries with initial ratios of $30 < a_B/R_B < 100$.}
\label{fig:SEMI_RUN2}
\end{figure}

Our second set of simulations, in which we integrate 360 binaries with $30 < a_B/R_B < 100$ through our first simulation set's planetary encounters, helps further test this possibility. Fig. \ref{fig:SEMI_RUN2} shows the final distribution of binary semimajor axes from these runs. Here we see that $\sim$10\% of surviving binaries finish the simulation with $a_B/R_B > 100$ with a tail to very large values of $a_B/R_B$, again supporting the idea that moderately tight primordial binaries generate a small population of widely spaced binaries in the modern HCB. 

When we account for binaries in the second simulation set disrupted by planetary encounters, we find that 5\% of binaries with initial separations of $30 < a_B/R_B < 100$ evolve to final separations of $a_B/R_B > 100$. This is comparable to the survival rate of primordial binaries with initial separations of $a_B/R_B > 100$ \citep{nesvorny2019binary}. Thus, a small number of very widely spaced binaries in the modern HCB should be expected from a modest population (5--10\% of primordial belt objects) of moderately spaced binaries in the primordial belt, which is already necessary to explain the modern HCB binaries with $30 < a_B/R_B < 100$. 

Due to our simulations' coarse resolution, collision modeling is beyond their scope, yet it may be important \citep{nesvorny2019binary}. If small impacting KBOs alter primordial binary separations, they could enhance the diffusion of tighter systems to wider ($a_B/R_B > 100$) systems and also further lower the survival chance of initially wide binaries \citep{parker2011collisional, bru16}. Alternatively, collisions may not enhance the diffusion of tighter binaries but instead increase catastrophic binary ionizations, but in this case, the effects would be even more severe for initially wide binaries. In this scenario, our primordial fraction of $30 < a_B/R_B < 100$ binaries would need to be higher, but the required primordial fraction of $a_B/R_B > 100$ would need to be higher still if diffusion of tighter binaries is not considered. Thus, we argue that wide binary formation via diffusion is still plausible when collisions are considered.

\begin{figure}
\includegraphics[scale=0.24]{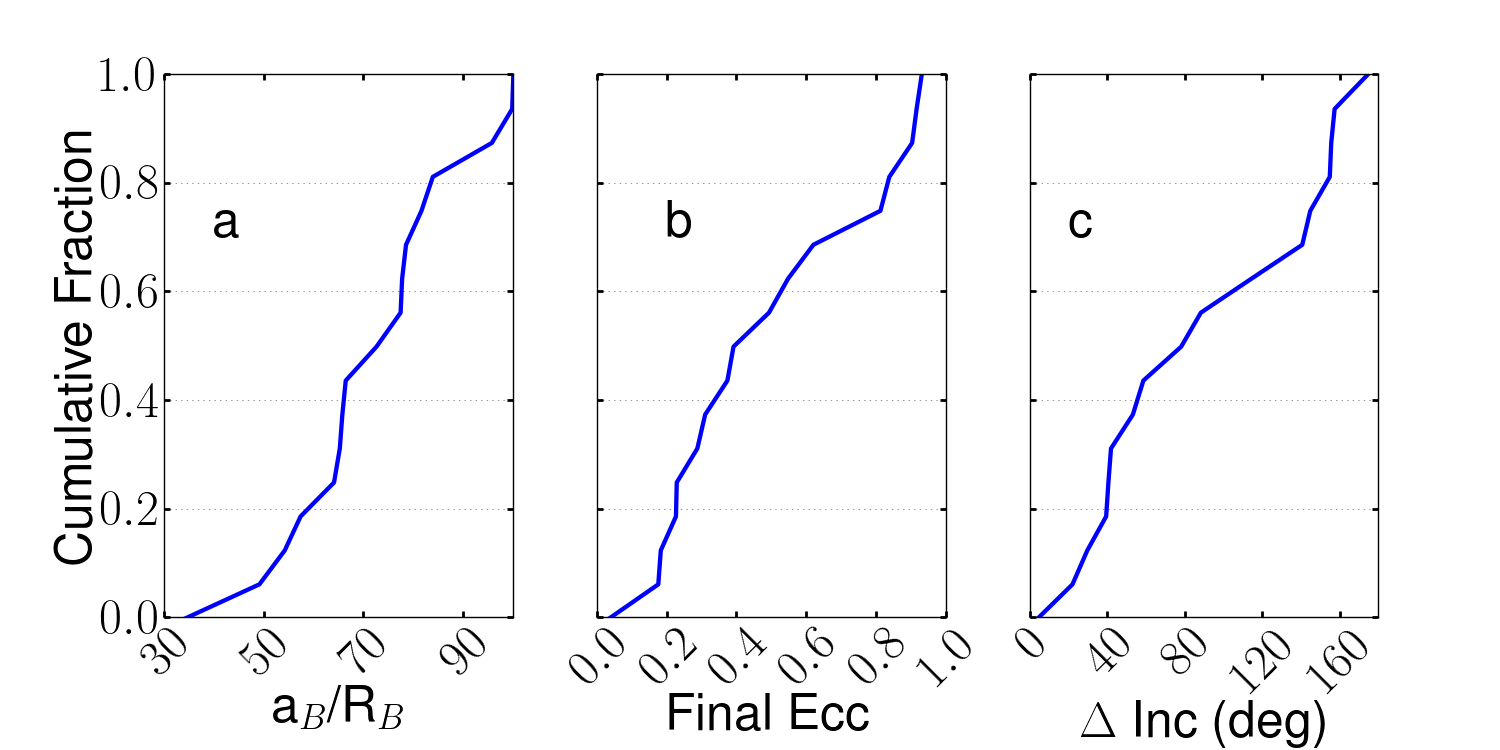}
\caption{Information is shown for the second simulation set's binaries that evolve from initial orbits of $30 < a_B/R_B < 100$ to final orbits of $a_B/R_B > 100$. Plot (a) shows the cumulative distribution of these binaries' initial ratios of semimajor axes to binary radii. Plot (b) shows the cumulative distribution of these binaries' final eccentricities. Plot (c) shows cumulative distribution of the angular change in orbital planes for these binaries.}
\label{fig:SEMI_RUN2_WB}
\end{figure}

In Fig. \ref{fig:SEMI_RUN2_WB} (a), we look at the initial separations of the binaries that evolved to separations of $a_B/R_B > 100$. While binaries that evolve to $a_B/R_B > 100$ are somewhat more likely to have initial separations near $a_B/R_B \sim 100$, these types of initial separations do not dominate our binary sample and all portions of our initial parameter space are capable of generating very wide binaries. 

In plots (b) and (c) of Fig. \ref{fig:SEMI_RUN2_WB}, we study the final distributions of eccentricities and inclinations for our binaries that evolved to separations of $a_B/R_B > 100$. In plot (b) we see this collection of initially nearly circular orbits has been dramatically transformed by planetary encounters. At the end of our simulations there is a very wide range of final eccentricities, with a median eccentricity near 0.4. Similarly, the orbital planes of these binaries are generally reoriented by large angles. In plot (c) we calculate the angle between each binary's initial and final angular momentum vectors and plot the distribution of such angles. We see that $\sim$60\% of the orbital planes are altered by at least 90$^{\circ}$. Thus, if these binaries had nearly coplanar orbits initially, we would predict they should now have a nearly isotropic distribution of inclinations. Thus, if the widest binaries in the HCB have an inclination distribution that is strongly concentrated toward coplanar orbits, this would likely exclude the possibility that they evolved from tighter primordial binaries.


 \section{Summary \& Conclusions}

 In line with previous work \citep{nesvorny2019binary, parker2010destruction}, we find that many binaries are destroyed by close encounters with giant planets during their implantation into the HCB. In particular, the survival rate drops off dramatically at initial binary separations of $a_B/R_B \sim 100$. As originally found by \citet{nesvorny2019binary}, our work suggests that the primordial belt would need a very large fraction ($\sim50\%$) of binaries with $a_B/R_B \gtrsim 100$ if these primordial binaries had to yield the small number of such binaries seen today.
 
 However, we also suggest an alternative explanation for this small number of very widely spaced binaries in the modern HCB. Our work shows that planetary encounters will soften $\sim$5\% of binaries with initial separations of $30 < a_B/R_B < 100$ into binaries with $a_B/R_B > 100$. It may be that the few very widely separated binaries observed in the modern HCB are derived from a larger population of somewhat tighter primordial binaries that were softened during the formation of the modern Kuiper belt. This primordial population of tighter binaries may generate the modest population of moderately spaced binaries in the modern HCB as well as the handful of very wide binaries in the modern HCB. 
 
 \section{Acknowledgements}
 
This work was performed with support from NASA Emerging Worlds grant
80NSSC18K0600. Our computing was performed at the OU Supercomputing Center for Education \& Research (OSCER) at the University of Oklahoma (OU).
 
 \section{Data availability}
The data underlying this article will be shared on reasonable request to the corresponding author.

\bibliographystyle{apj}
\bibliography{Bibliography.bib}

\end{document}